\documentclass[aps,prd,two column,showpacs,preprintnumbers,amsmath,amssymb]{revtex4}

\usepackage{graphicx}
\usepackage{dcolumn}
\usepackage{bm,epsfig}

\begin{document}

\title{A study of $S$-wave $DK$ interactions in the chiral SU(3) quark model}

\author{ZHANG Dan£¨Õŵ¤£©}
\email{zhangdan77@gmail.com; Tel:13947105425}
\author{ZHAO Qiao-Yan£¨ÕÔÇÉÑࣩ}
\affiliation{
Department of physics, Inner Mongolia University, Hohhot 010021, China}
\author{ZHANG Qiu-Yang£¨ÕÅÇïÑô£©}
\affiliation{
College of Information $\&$ Electronic Engineering, Zhejiang GongShang University, Hangzhou 310018, China}


\begin{abstract}

The $DK$ interaction is relevant to the interpretation of the
$D_{sJ}(2317)$. We dynamically investigate $S$-wave $DK$
interactions in the chiral SU(3) quark model by solving the
resonating group method equation. The numerical results show an
attraction between $D$ and $K$, which is from boson exchanges
between light quarks. However, such an attraction is not strong
enough to form a $DK$ molecule. Meanwhile, $S$ partial wave phase
shifts of $DK$ elastic scattering are obtained. The case of $S$-wave
$D^*K$ is rather similar to that of $DK$. To draw a definite
conclusion whether a molecular state exists in $DK$ or $D^*K$
system, more details of dynamics should be considered in further
study.

\end{abstract}


\maketitle

In 2003 BaBar collaboration reported a narrow positive-parity scalar
meson $D_{sJ}(2317)$ [1], which was confirmed by CLEO later [2]. In
the same experiment CLEO observed the $1^+$ partner state at $2460$
MeV [2]. These two states lie below $DK$ and $D^*K$ thresholds,
respectively. So far there have been lots of experimental
investigations of these two narrow resonances [3]. Their results are
consistent with the spin-parity assignments of $J^P=0^{+}$ for
$D_{sJ}(2317)$ and $J^P=1^{+}$ for $D_{sJ}(2460)$.

The discovery of these two states has triggered heated discussion on
their nature [4], and the key point is how to interpret their low
masses. It's tempting to explain these two states as $(0^+,1^+)$
$P$-wave $c\bar s$ doublet [5], tetraquark states [6,7], or the
$c\bar s(0^+,1^+)$ spin parity partners of the $(0^-,1^-)$ doublet
in the framework of chiral symmetry [8]. However, the predicted
masses in quark model or in lattice QCD calculation [9] are higher
than the low experimental data. In addition, it is pointed out [10]
that $D_{sJ}(2317)$ might receive a large component of $DK$. From
the experience with $a_0/f_0(980)$, the low mass of $D_{sJ}(2317)$
could arise from the mixing between the $0^+$ $c\bar s$ state and
the $DK$ continuum [11]. $D_{sJ}(2317)$ is also proposed to be a
dominantly $I=0$ $DK$ state with some $I=1$ admixture [12].

As mentioned above, it is worthwhile to study $DK$ interactions
dynamically with various methods to further understand the nature of
the $D_{sJ}(2317)$. In this paper, we will investigate $S$-wave $DK$
interactions in the chiral SU(3) quark model by solving the
resonating group method(RGM) equation [13].

The chiral SU(3) quark model [14] is a useful tool in connecting the
QCD theory and the experimental observables, especially for the
light quark systems. In this model, the quark-quark interaction
contains confinement, one gluon exchange (OGE) and pseudoscalar and
scalar meson exchanges. It has been proved successful in reproducing
the energies of the baryon ground states, the binding energy of the
deuteron, the nucleon-nucleon($NN$) scattering phases and the
hyperon-nucleon($YN$) cross sections. Recently the chiral SU(3)
quark model has been extended to study the baryon-meson interactions
[15], baryon-antibaryon system [16], and states including heavy
quarks [7,17]. In the present letter, we will follow the methods in
above works to study $DK$ systems.

The paper is organized as follows. Firstly, we briefly describe the
theoretical frame including the model hamiltonian and parameters.
Then numerical results are shown and discussed, and the summary is
presented finally.

The chiral SU(3) quark model has been widely described in the
literature [7,14-17] and we just give its salient feature here. The
Hamiltonian of the $DK$ system can be written as
\begin{eqnarray}
H=\sum_iT_i-T_G+\sum_{i<j=1}^4V_{ij}\; ,
\end{eqnarray}
where $T_G$ is the kinetic energy operator for the c.m. motion, and
$V_{ij}$ represents the interactions between $qq$, $q\bar q$ or
$\bar q\bar q$.

As for $qq$ pair,
\begin{eqnarray}
V_{qq}(ij)=V_{qq}^{conf}+V_{qq}^{OGE}+V_{qq}^{ch},
\end{eqnarray}
where the confinement potential $V_{qq}^{conf}$ is taken as linear
form in this work. $V_{qq}^{ch}$ represents the interaction from
chiral field coupling, which includes scalar and pseudoscalar boson
exchanges in the chiral SU(3) quark model,
\begin{eqnarray}
V^{ch}_{qq}(ij) = \sum_{a=0}^8 V_{\sigma_a}({\bm
r}_{ij})+\sum_{a=0}^8 V_{\pi_a}({\bm r}_{ij}),
\end{eqnarray}
where $\sigma_{0},...,\sigma_{8}$ are the scalar nonet fields, and
$\pi_{0},..,\pi_{8}$ the pseudoscalar nonet fields.

Replacing the color part $(\lambda^c_i\cdot\lambda^c_j)$ in
$V_{qq}^{conf}$ and $V_{qq}^{OGE}$ by
($\lambda^{c\ast}_i\cdot\lambda^{c\ast}_j)$, we can obtain $V_{\bar
q\bar q}^{OGE}$ and $V_{\bar q\bar q}^{conf}$. $V_{\bar q\bar
q}^{ch}$ has the same form as $V_{qq}^{ch}$.

The interaction of $q\bar q $ pair includes two parts: direct
interaction and annihilation part
\begin{eqnarray}
V_{q\bar q}=V_{q\bar q}^{dir}+V_{q\bar q}^{ann}\;  ,
\end{eqnarray}
\begin{eqnarray}
V_{q\bar q}^{dir}=V_{q\bar q}^{conf}+V_{q\bar q}^{OGE}+V_{q\bar
q}^{ch}\; .
\end{eqnarray}
For a preliminary investigation, we neglect the contribution of
annihilation part in the present work. $V_{q\bar q}^{dir}$ can be
obtained from $V_{qq}$. As for $V_{q\bar q}^{conf}$ and $V_{q\bar
q}^{OGE}$, the transformation from $V_{qq}$ to $V_{q\bar q}$ is
given by $\lambda _i^c\cdot \lambda _j^c\rightarrow -\lambda
_i^c\cdot \lambda _j^{c*}$, while
\begin{equation}
V_{q\bar{q}}^{ch}=\sum_{j}(-1)^{G_j}V_{qq}^{ch,j}.
\end{equation}
Here $(-1)^{G_j}$ represents the G parity of the $j$th meson. The
detailed expressions can be found in Refs. [7,14-17].

Note that for the heavy-light quark pairs, the Goldstone boson
exchanges will not be considered as a primary study. We use the same
cutoff $\Lambda$ for various mesons. Its value is around the scale
of chiral symmetry breaking($\sim 1$ GeV).

\begin{table}[htb]
\centering \caption{Model parameters for the light quark pairs. The
meson masses are: $m_{\sigma^\prime}=m_{\epsilon}=m_{\kappa}=980$
MeV, $m_\pi=138$ MeV, $m_K=495$ MeV , $m_\eta=549$ MeV,
$m_{\eta^\prime}=957$ MeV.}\label{paras}
\setlength{\tabcolsep}{2.6mm}
\begin{tabular}{cccccccc}
\hline\hline
\multicolumn{8}{l}{Model Parameters}\\
\hline
$b_u$ (fm)& & & & & & &0.5 \\
$m_u$ (MeV)& & & & & & & 313\\
$m_s$ (MeV)& & & & & & & 470\\
$g_{ch}$ & & & & & & & 2.621 \\
$m_\sigma$ (MeV)& & & & & & & 595\\
$\theta^{ps}$& & & & & & & -23$^{\circ}$\\
$\theta^{s}$& & & & & & & 0$^{\circ}$\\
\hline\hline
\end{tabular}
\end{table}

The parameters for the light quark pairs are taken from the previous
work [14], which can give a satisfactory description of the energies
of the baryon ground states, the binding energy of deuteron, the
$NN$ scattering phase shifts, and $NY$ cross sections. For
simplicity, we only show them as Table \ref{paras}, where the
harmonic-oscillator width parameter $b_u=0.50$ fm. The up (down)
quark mass $m_{u(d)}$ and the strange quark mass $m_s$ are taken to
be the usual values: $m_{u(d)}=313$ MeV and $m_s=470$ MeV. The
coupling constant for scalar and pseudoscalar chiral field coupling,
$g_{ch}$, is determined according to the relation
\begin{eqnarray}
\frac{g^{2}_{ch}}{4\pi} = \left( \frac{3}{5} \right)^{2}
\frac{g^{2}_{NN\pi}}{4\pi} \frac{m^{2}_{u}}{M^{2}_{N}},
\end{eqnarray}
with empirical value $g^{2}_{NN\pi}/4\pi=13.67$. The mass of the
phenomenological $\sigma$ meson is treated as an adjustable
parameter, and we take $m_{\sigma}=595$ MeV in the chiral SU(3)
quark model. For other meson masses, we use the experimental values.
$\eta, \eta^{\prime}$ mesons are mixed by $\eta_1, \eta_8$
\begin{eqnarray}
\eta=\eta_8 \cos \theta^{ps}-\eta_1 \sin \theta^{ps},\nonumber\\
\eta^{\prime}=\eta_8 \sin \theta^{ps}+\eta_1 \cos \theta^{ps},
\end{eqnarray}
and the mixing angle $\theta^{ps}$ is taken to be the usual value
with $\theta^{ps}=-23^{0}$. Usually $\sigma, \epsilon$ mesons are
mixed by $\sigma_1, \sigma_8$
\begin{eqnarray}
\sigma=\sigma_8 \sin \theta^{s}+\sigma_1 \cos \theta^{s},\nonumber\\
\epsilon=\sigma_8 \cos \theta^{s}-\sigma_1 \sin \theta^{s}.
\end{eqnarray}
The mixing angle $\theta^s$ is an open problem because the structure
of the $\sigma$ meson is unclear and controversial. Firstly the
scalar meson mixing is not considered, i.e. $\theta^s=0^{\circ}$.

To investigate the heavy quark mass dependence, we take several
typical values $m_c=1430$ MeV [7], $m_c=1550$ MeV [18], $m_c=1870$
MeV [19].

The OGE coupling constants and the confinement strengths can be
derived from the masses of ground state baryons and heavy mesons
[7,14]. Between the two color-singlet clusters $D(c\bar u)$ and
$K(u\bar s)$, there is no OGE interaction and the confinement
potential scarcely contributes any interaction. Therefore these
values will not affect the final results and we do not present them
here.

To explore the effect of the cutoff, we use two values
$\Lambda=1100$ MeV and $\Lambda=1500$ MeV.

With the parameters determined, the $S$-wave $DK$ system can be
dynamically studied in the framework of the RGM, a well established
method for detecting the interaction between two clusters. The
details of solving the RGM equation can be found in Refs.
[13,15,16]. By solving the RGM equation, one gets the energy of the
system, the relative motion wave function, and the elastic
scattering phase shifts.

As mentioned above, in our present study two parts are not
considered: (1) the chiral field induced interactions between heavy
and light quarks; (2) the $s$-channel annihilation interactions
between $q$ and $\bar q$. In addition, in the chiral SU(3) quark
model, only scalar and pseudoscalar meson exchanges are involved.

\begin{figure}[ht]
\epsfig{file=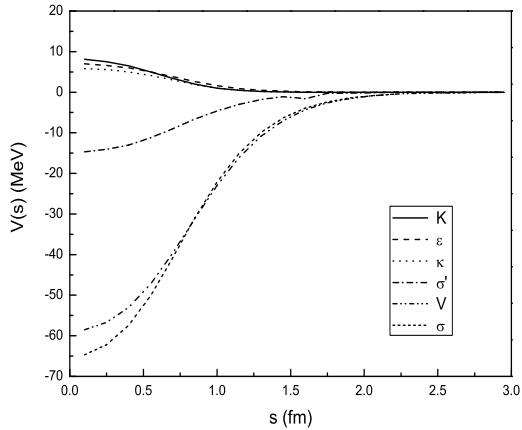,width=8.0cm,height=7.0cm} \vglue -0.5cm
\caption{\small \label{vs} The effective potential V(s) for
different meson exchange, which is independent of $m_c$. The
parameters are taken from Table \ref{paras}, and $\Lambda=1100$ MeV.
From top to bottom, the curves correspond to the contributions from
$K, \epsilon, \kappa, \sigma^\prime$ mesons, total contribution of
all mesons $V$, and $\sigma$ meson.}
\end{figure}

Firstly, we apply the RGM calculation to the $S$-wave $DK$ isospin
$I=0$ system. Before the numerical evaluation, let's take a look at
the effective potential
\begin{equation}
V(s)=V^{L=0}(s,s),
\end{equation}
where the generator coordinate $s$ can qualitatively describe the
distance between the two clusters. The potentials corresponding to
various considerations with $\Lambda=1100$ MeV are illustrated in
Fig. \ref{vs}. From this figure, we can see that the total potential
$V(s)$(Line '$V$' in Fig. \ref{vs}) is attractive, which is relies
on meson exchanges. $\sigma$ and $\sigma^{\prime}$ mesons provide
considerable attractions, while the interactions due to $K,
\epsilon, \kappa$ are weakly repulsive with comparable amplitudes.
And $\pi, \eta, \eta^{\prime}$ have no contribution.

The further numerical calculation shows that all potentials are
independent of the mass of $c$ quark $m_c$. It is obvious since we
do not consider meson exchanges between $c$ and light quarks. The
results with $\Lambda=1500$ MeV are similar to those demonstrated in
Fig. \ref{vs}, but amplitudes of all curves are a little bigger,
which lead to the attractive potential $V$ at most $4-6$ MeV
stronger.

\begin{figure}[ht]
\epsfig{file=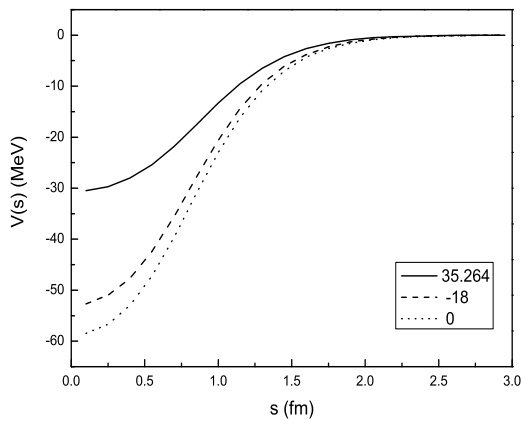,width=8.0cm,height=7.0cm} \vglue -0.5cm
\caption{\small \label{vcom} The total effective potential V(s) with
different $\theta^s$. The parameters are taken from Table
\ref{paras}, except for $\theta^s$, and $\Lambda=1100$ MeV. From top
to bottom, the curves correspond to $\theta^s=35.264^{\circ},
\theta^s=-18^{\circ}, \theta^s=0^{\circ}$.}
\end{figure}

\begin{figure}[htb]
\epsfig{file=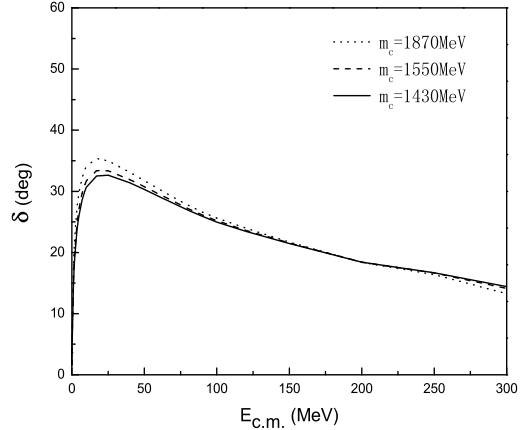,width=8.0cm,height=7.0cm} \vglue -0.5cm
\caption{\small \label{phase} The $S$-wave $DK$ elastic scattering
phase shifts with $\Lambda=1100$ MeV. From top to bottom, the curves
correspond to $m_c$ being $1430, 1550$ and $1870$ MeV.}
\end{figure}

To present the effect of the mixing angle, we take two more values
$\theta^s=35.264^{\circ}$ and $\theta^s=-18^{\circ}$ [20]. As shown
in Fig \ref{vcom}, the attractions become weaker when $\theta^s \neq
0^{\circ}$. On the other hand, we find the $DK$ effective potential
is independent of the pseudoscalar mesons mixing angle
$\theta^{ps}$, since $\eta, \eta^{\prime}$ mesons have no
contribution(as shown in Fig \ref{vs}).

Then it is natural for us to wonder whether such an attraction can
form a $DK$ bound state. The binding energy of the $DK$ system [21]
is calculated by solving RGM equation. After exploring all possible
combinations of the parameters in the former section, we fail to get
a bound state of $DK$.

In order to get more information, we study the $DK$ elastic
scattering processes, and the phase shifts of $S$ partial waves are
shown in Fig. \ref{phase}, where $\Lambda=1100$ MeV. This figure
indicates that the interactions are weakly attractive in the middle
energy range. When we take $\Lambda=1500$ MeV, the curves are a
little higher shifted,
implying the attractions are a little stronger. We also find that
the mass of $c$ quark gives little contribution. The analysis of $S$
partial waves phase shifts is qualitatively consistent with that of
$V(s)$.

All results we have presented above are based on $DK$ $I=0$. Our
further calculations suggest that the results are isospin
independent, i.e., the case of $I=1$ are the same as that of $I=0$.
A part reason is $\sigma$ exchange plays dominant role in the $DK$
interactions, which is isospin independent. Another possible reason
is meson exchanges between $c$ and light quarks are not included.

In addition, we perform the same calculation to the $S$-wave $D^* K$ system,
and the rather similar results are obtained.

In this work we have dynamically studied the interactions of
$S$-wave $DK$ system by solving RGM equation in the chiral SU(3)
quark model, including bound state problem and elastic scattering
phase shifts. We have obtained some useful information. In our
present calculation the potentials between $D$ and $K$ two clusters
come from meson exchanges. By taking parameters shown in Table
\ref{paras}, We find the attractions provided by $\sigma$ and
$\sigma^\prime$ are stronger than the repulsions from
$K,\epsilon,\kappa$, which results in $DK$ interaction is
attractive. However, such an attraction is not strong enough to form
a $DK$ bound state. Moreover, the values of $\Lambda$ and $m_c$
offer little help to the $DK$ interaction, the scalar mesons mixing
angle $\theta^s \neq 0$ can weaken the attractions, and the results
are independent of isospin. The information extracted from the $S$
partial phase shift of $DK$ is qualitatively consist with that of
bound state problem. Additionally, the case of $S$-wave $D^* K$ is
rather similar to that of $S$-wave $DK$ system.

In order to determine the nature of $DK$ interactions and whether
$DK$ or $D^* K$ molecule exists, we will take detailed study in
future, including: (1)to involve the vector meson exchanges, which
are expected to contribute more attractions; (2)to consider the
chiral field induced interactions between heavy and light quarks;
(3)to study the $s$-channel annihilation interactions between $q$
and $\bar q$.

\section*{Acknowledgments}

Dan Zhang thanks Professor Zong-Ye Zhang for helpful discussions.
This project was supported by the National
Natural Science Foundation of China under Grants 10847144, the Inner
Monglia Education Foundation (NJzy08006), the Inner Monglia University
Foundation,and the Inner Monglia University Postdoctoral Science
Foundation.


\begin{thebibliography}{99}
\bibitem{bind}The binding energy of $DK$ system is definited as $E_0=M_{DK}-M_{D}-M_{K}$. If $E_0$ is negative,
              the system is bound.
\end{thebibliography}
\end{document}